\documentclass{aa}
\usepackage{epsfig,amsfonts,amssymb}
\def\lsim{{\lesssim}}

\def\eros{{\sc eros}}
\def\macho{{\sc macho}}
\def\ogle{{\sc ogle}}
\def\lmc{{\sc lmc}}
\def\smc{{\sc smc}}
\def\ie{{\em i.e.}}

\begin{document}

\thesaurus{10.08.1, 10.11.1, 10.19.2, 11.13.1, 12.04.1, 12.07.1}

\title{Not enough stellar Mass Machos in the Galactic Halo
\thanks{Based on observations made at the European Southern Observatory,
La Silla, Chile.}}
\author{
T.~Lasserre\inst{1},
C.~Afonso\inst{1},
J.N.~Albert\inst{2},
J.~Andersen\inst{6},
R.~Ansari\inst{2}, 
\'E.~Aubourg\inst{1}, 
P.~Bareyre\inst{1,4}, 
F.~Bauer\inst{1},
J.P.~Beaulieu\inst{3},
G.~Blanc\inst{1},
A.~Bouquet\inst{4},
S.~Char$^{\dag}$\inst{7},
X.~Charlot\inst{1},
F.~Couchot\inst{2}, 
C.~Coutures\inst{1}, 
F.~Derue\inst{2}, 
R.~Ferlet\inst{3},
J.F.~Glicenstein\inst{1},
B.~Goldman\inst{1},
A.~Gould\inst{8,1\,}\thanks{Alfred P.\ Sloan Foundation Fellow},
D.~Graff\,\inst{8,1},
M.~Gros\inst{1}, 
J.~Ha\"{\i}ssinski\inst{2}, 
J.C.~Hamilton\inst{4},
D.~Hardin\inst{1},
J.~de Kat\inst{1}, 
A.~Kim\inst{4},
\'E.~Lesquoy\inst{1,3},
C.~Loup\inst{3},
C.~Magneville \inst{1}, 
B.~Mansoux\inst{2}, 
J.B.~Marquette\inst{3},
\'E.~Maurice\inst{5}, 
A.~Milsztajn \inst{1},  
M.~Moniez\inst{2},
N.~Palanque-Delabrouille\inst{1}, 
O.~Perdereau\inst{2},
L.~Pr\'evot\inst{5}, 
N.~Regnault\inst{2},
J.~Rich\inst{1}, 
M.~Spiro\inst{1},
A.~Vidal-Madjar\inst{3},
L.~Vigroux\inst{1},
S.~Zylberajch\inst{1}
\\   \indent   \indent
The EROS collaboration
}
%
\institute{
CEA, DSM, DAPNIA,
Centre d'\'Etudes de Saclay, F-91191 Gif-sur-Yvette Cedex, France
\and
Laboratoire de l'Acc\'{e}l\'{e}rateur Lin\'{e}aire,
IN2P3 CNRS, Universit\'e de Paris-Sud, F-91405 Orsay Cedex, France
\and
Institut d'Astrophysique de Paris, INSU CNRS,
98~bis Boulevard Arago, F-75014 Paris, France
\and
Coll\`ege de France, Physique Corpusculaire et Cosmologie, IN2P3 CNRS, 
11 pl. M. Berthelot, F-75231 Paris Cedex, France
\and
Observatoire de Marseille,
2 pl. Le Verrier, F-13248 Marseille Cedex 04, France
\and
Astronomical Observatory, Copenhagen University, Juliane Maries Vej 30, 
DK-2100 Copenhagen, Denmark
\and
Universidad de la Serena, Facultad de Ciencias, Departamento de Fisica,
Casilla 554, La Serena, Chile
\and
Departments of Astronomy and Physics, Ohio State University, Columbus, 
OH 43210, U.S.A.
}
\offprints{Thierry.Lasserre@cea.fr}

\date{Received 11 February 2000 / Accepted 23 February 2000}

\authorrunning{T. Lasserre et al.}
\titlerunning{Not enough stellar Mass \macho s in the Galactic Halo }

\maketitle

\begin{abstract}

We combine new results from the search for microlensing
towards the Large Magellanic Cloud (\lmc) by \eros2 
(Exp\'erience de Recherche d'Objets Sombres)
with limits previously reported by \eros1 and \eros2 
towards both Magellanic Clouds.  The derived upper limit on the 
abundance of stellar mass \macho s rules out such objects 
as an important component of the Galactic
halo if their mass is 
smaller than  $1 {\rm M}_{\odot}$. 
\keywords {Galaxy: halo -- Galaxy: kinematics and dynamics -- 
Galaxy: stellar content -- Magellanic Clouds -- dark matter -- 
gravitational lensing
}
\end{abstract}

\section{Research context}
        The search for gravitational microlensing in our Galaxy
has been going on for a decade, following the 
proposal to use this effect as a tool to probe the dark matter content
of the Galactic halo (\cite{pac86}).  The first microlensing 
candidates were reported in 1993, towards the \lmc\
(\cite{aub93}; Alcock et al. 1993) and the Galactic Centre (\cite{uda93})
by the \eros , \macho\ and \ogle\ collaborations.

Because they observed no microlensing candidate with a duration
shorter than 10~days,
the \eros1 and \macho\ groups were able to exclude 
the possibility that a substantial 
fraction of the Galactic dark matter resides in planet-sized objects
(\cite{aub95}; \cite{alc96}; Renault et al. 1997; 
\cite{ren98}; \cite{alc98}).  

However a few events were
detected with longer time\-scales.  From 6-8 candidate events
towards the \lmc , the \macho\ group estimated an optical depth of
order half that required to account for the dynamical mass of the 
standard spherical dark halo;  
the typical Einstein radius crossing time of the events, $t_E$, 
implied an average mass of about 0.5~M$_\odot$ for the lenses 
(Alcock et al. 1997a).
Based on two candidates, \eros1 set an upper limit on the 
halo mass fraction in objects of similar masses
(Ansari et al. 1996), that is below that required to 
explain the rotation curve of our Galaxy\footnote{
Assuming the \eros1 candidates are microlensing events, 
they would correspond to an optical depth six times lower than that
expected from a halo fully comprised of \macho s.}.

The second phase of the \eros\ programme was started in 1996, with a 
ten-fold increase in
the number of monitored stars in the Magellanic Clouds.
The analysis of the first two years of data towards the 
Small Magellanic Cloud (\smc)
allowed the detection of one microlensing event 
(Palanque-Delabrouille et al. 1998; see also Alcock et al., 1997b).
This single event, out of 5.3 million stars, allowed \eros2 
to further constrain the halo composition, excluding in 
particular that more than 50~\% of the standard dark halo
is made up of $0.01 - 0.5 \:{\rm M}_\odot$ objects 
(Afonso et al. 1999).  
In contrast, an optical detection of a halo
white dwarf population was reported (Ibata et al. 1999).

In this letter, we describe the analysis of the two-year light curves
from 17.5 million \lmc\ stars.  We then 
combine these results with our previous limits, and derive the
strongest limit obtained thus far on the amount of stellar mass
objects in the Galactic halo.

\section{Experimental setup and LMC observations}
	The telescope, camera, telescope operation and data reduction
are as described in Bauer et al. (1997) and 
Pa\-lan\-que-Delabrouille et al. (1998). 
Since August 1996, we have been monitoring
66 one square-degree fields in the \lmc , 
simultaneously in two wide passbands. 
Of these, data prior to May 1998 from 25~square-degrees 
spread over 43~fields have been analysed.  In this period, about
70-110~images of each field were taken, with exposure times ranging from 
3~min in the \lmc\ center to 12~min on the periphery; 
the average sampling is once every 6~days. 

\section{LMC data analysis}
        The analysis of the \lmc\ data set was done using a program
independent from that used in the \smc\ study, with largely
different selection criteria.
The aim is to cross-validate both programs 
(as was already done in the analysis
of \eros1 Schmidt photographic plates, \cite{ans96})
and avoid losing rare microlensing events.
Preliminary results of the present analysis were reported in 
Lasserre (1999).
We only give here a list of the various steps, as well as a
short description of the principal new features; 
details will be provided in Lasserre et al. (2000).

We first select the 8\% ``most variable'' light curves, a sample
much larger than the number of detectable variable stars.
Working from this ``enriched'' subset, we apply a first 
set of cuts to select, in each colour separately, 
the light curves that exhibit significant variations.
We first identify the baseline flux in the light curve - basically
the most probable flux.  
We then search for {\it runs} along the light curve,
\ie\ groups of consecutive measurements that are all on the same side 
of the baseline flux.
We select light curves that either
have an abnormally low
number of runs over the whole light curve, or
show one long run (at least 
5 valid measurements) that is very 
unlikely to be a statistical fluctuation. 
We then ask for a minimum signal-to-noise ratio by requiring that
the group of 5 most luminous
consecutive measurements be significantly further
from the baseline than the average spread of the measurements.
We also check that the measurements inside the most significant run
show a smooth time variation.

The second set of cuts compares the measurements with the best fit
point-lens point-source constant speed microlensing light curve
(hereafter ``simple microlensing'').  They
allow us to reject variable stars whose light curve differs too much
from simple microlensing, and are sufficiently loose not to reject 
light curves affected by blending, parallax 
or the finite size of the source, 
and most cases of multiple lenses or sources. 

After this second set of cuts, stars selected 
in at least one passband represent
about 0.01\% of the initial sample; almost all of them
are found in two thinly populated zones of the colour-magnitude 
diagram. The third set of cuts deals with this physical background.
The first zone
contains stars brighter and much redder than those of the red clump;
variable stars in this zone are rejected if they vary by less than
a factor two or have a very poor fit to simple microlensing.  
The second zone is the top of the main sequence.  Here we find that
selected stars, known as blue bumpers (\cite{alc97a}), 
display variations that are
always smaller than 60\% and lower in the visible passband 
than in the red one.  These cannot correspond to simple microlensing,
which is achromatic; they cannot correspond to microlensing
plus blending with another unmagnified star either, 
as it would imply blending by even bluer stars,
which is very unlikely. 
We thus reject all candidates from the second zone
exhibiting these two features.

The fourth set of cuts tests for compatibility between the
light curves in both passbands.
We retain candidates selected in only one passband
if they have no conflicting data in the other passband.
For candidates selected independently in the two passbands,
we require that their largest variations coincide in time.

The tuning of each cut and the calculation of the microlensing 
detection efficiency are done with 
simulated simple microlensing light curves, as described in 
Pa\-lanque-Dela\-brouille et al. (1998). 
For the efficiency calculation, 
microlensing parameters are drawn uniformly in the following
intervals: time of maximum magnification $t_0$ 
within the observing period $\pm 150$~days,
impact parameter normalised to the Einstein
radius $u_0 \in [0,2]$ and timescale 
$t_E \in [5,300]$ days.  
All cuts on the data were also applied to the simulated
light curves.

%
%
\begin{figure} [ht] 
 \begin{center} \epsfig{file=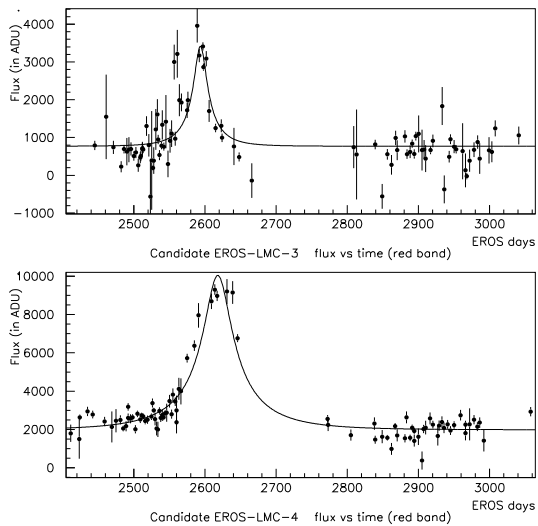,width=7.8cm} 
  \caption{Light curves of candidates EROS-LMC-3 and 4. 
   The plain curves show the best point-lens point-source fits;
   time is in days since Jan. 1, 1990 (JD 2,447,892.5).}  
  \label{cdl_evts}
 \end{center} 
\end{figure}
%
%

Only two candidates remain after all cuts. 
Their light curves are shown
in Fig.~\ref{cdl_evts}; microlensing fit parameters
are given in Table~\ref{eventparm}. 
Although the candidates pass all cuts, agreement with simple 
microlensing is not excellent.

%
%
\begin{table}[ht]
\begin{center} \vspace{-0.0cm} 
\begin{tabular}{|l||c|c|c|c|c|c|c|}
\hline
   &$u_0$&$t_E$&$c_{\rm \,bl}^R$&$c_{\rm \,bl}^V$&$\chi^2/{\rm dof}$& 
$V_J$&$R_C$ \\
\hline
 \lmc-3 & $0.23$&$41$&$0.76$& 1 &208/145&22.4&21.8\\
\hline
 \lmc-4 & $0.20$&$106$& 1 & 1 &406/150&19.7&19.4\\
\hline

\end{tabular}
\caption{Results of microlensing fits to the two new \lmc\ candidates;
 $t_E$ is the Einstein radius crossing time in days, 
$u_{0}$ the impact parameter, and $c_{\rm \,bl}^{R(V)}$ the 
$R(V)$ blending coefficients.
}
\label{eventparm}
\end{center} \vspace{-0.5cm}
\end{table}
%
%

The efficiency of the analysis, normalised to events with an impact 
parameter $u_0<1$ and to an observing period $T_{\rm obs}$ of two
years, is summarised in Table~\ref{eff}.
The main source of systematic error is the uncertainty in the 
influence of blending. Blending lowers the observed magnifications
and timescales.  While this decreases the efficiency for a given star, 
the effective number of monitored stars is increased so that there
is partial compensation.
This effect was studied with synthetic images using measured magnitude
distributions (\cite{pal97}). 
Our final efficiency is within 10\% of the naive efficiency.
%
%
\begin{table}[ht]
\vspace{-0.3cm}
 \begin{tabular}{|c||c|c|c|c|c|c|c|c|c|c|} \hline
$t_E$ & 5 & 11& 18& 28& 45& 71& 112& 180& 225& 280\\ \hline
$\epsilon$ & 2 &  5& 11& 15& 19& 23&  26&  25&  18& 2.5\\ \hline
\end{tabular}
\caption{Detection efficiency in \% as a function of the
Einstein radius crossing time $t_E$ in days, 
normalised to events generated with $u_0<1$, and
to $T_{\rm obs}=2{\rm \: yrs}$. 
}
\label{eff} 
\vspace{-0.6cm}
\end{table}
%
%

\section{EROS1 results revisited}
     The two \eros1 microlensing candidates have been monitored by 
\eros2. 
The source star in event \eros -\lmc -2 had been found to be variable 
(\cite{ans95}), but microlensing fits taking into account 
the observed periodicity
gave a good description of the measurements.  Its follow-up by \eros2
revealed a new bump in March 1999, eight years after the first 
one\footnote{We thank the \macho\ group for communication about 
their data on this star.}.
This new variation, of about a factor two, was not well sampled 
but is significant.
Therefore, \eros -\lmc -2 is no longer a candidate
and we do not include it in the limit computation.

\section{Limits on Galactic halo MACHOs}
    \eros\ has observed four microlensing candidates towards the
Magellanic Clouds, one from \eros1 and two from \eros2 towards
the \lmc , and one towards the \smc .
As discussed in Palanque-Dela\-brouille et al. (1998),
we consider that the long duration
of the \smc\ candidate together with the absence of any detectable
parallax, in our data as well as in that of the \macho\ group
(\cite{alc97b}), indicates that it is most likely
due to a lens in the \smc .  For that reason, the limit
derived below uses the three \lmc\ candidates; for completeness,
we also give the limit corresponding to all four candidates.
 
The limits on the contribution of dark compact objects
to the Galactic halo are obtained by comparing the number 
and durations of 
microlensing candidates with those expected from Galactic
halo models.   
We use here the so-called ``standard'' halo model described in
Palanque-Delabrouille et al. (1998) as model 1.
The model predictions are computed for each \eros\
data set in turn, taking into account the corresponding detection 
efficiencies
(\cite{ans96}; Renault et al. 1998; 
Afonso et al. 1999; Table~\ref{eff} above),
and the four predictions are then summed.
In this model, all dark objects have the same mass $M$; we have
computed the model predictions for many trial masses $M$ in turn,
in the range [$10^{-8}\:{\rm M}_\odot$, $10^2\:{\rm M}_{\odot}$].

The method used to compute the limit is as  
in Ansari et al. (1996). We consider two ranges of
timescale $t_E$, smaller or larger than 
$t_E^{\rm lim} = 10$~days.  (All candidates have 
$t_E > t_E^{\rm lim}$.)
We can then compute, for each mass
$M$ and any halo fraction $f$,
the combined Poisson probability for obtaining,
in the four different \eros\ data sets taken as a whole, 
zero candidate with $t_E < t_E^{\rm lim}$ and 
three or less (alt.~four or less)
with $t_E > t_E^{\rm lim}$.  For any value of $M$,
the limit $f_{\rm max}$ is the value of $f$ for which this probability
is 5\%.  Whereas the actual limit depends somewhat on the precise
choice of $t_E^{\rm lim}$, the difference ($ \lsim 5\%$) is noticeable 
only for masses around $0.02\:{\rm M}_{\odot}$. 
Furthermore, we consider 10~days to be a conservative choice.
%
%
\begin{figure} [ht] 
 \begin{center} \epsfig{file=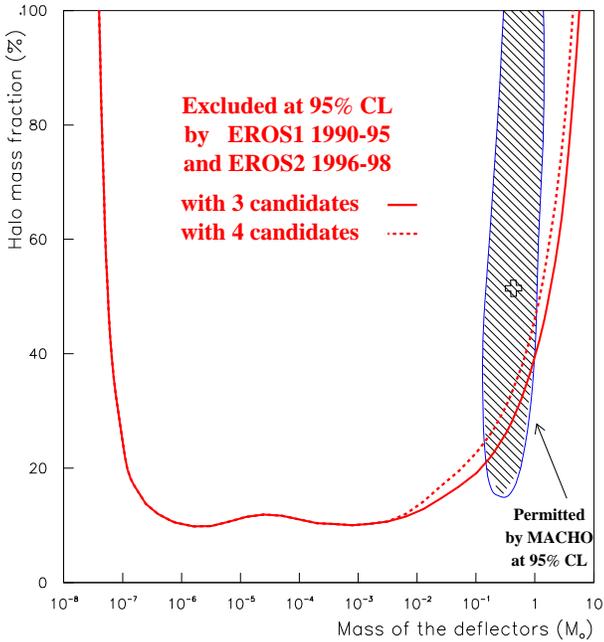,width=8.7cm}
  \vspace{-.1cm} \caption{ 95\% C.L. exclusion diagram on the halo mass
  fraction in the form of compact objects of mass $M$, for the
  standard halo model ($4 \times 10^{11}\:{\rm M}_\odot$ inside
  50~kpc), from all \lmc\ and \smc\ \eros\ data 1990-98.  
  The solid line is the limit inferred from the three \lmc\ microlensing 
  candidates; the dashed line includes in addition the \smc\
  candidate.  The {\sc macho} 95\%
  C.L. accepted region is the hatched area, with the preferred value
  indicated by the cross (Alcock et al. 1997a).} 
  \label{excl}
  \end{center} \vspace{-0.5cm}
\end{figure}
%
%

Figure \ref{excl} shows the 95\% C.L. exclusion limit derived 
from this analysis on the halo mass fraction, $f$,
for any given dark object mass, $M$. 
The solid line corresponds to the three \lmc\ candidates;
it is the main result of this letter. 
(The dashed line includes the \smc\ candidate in addition.)
This limit rules out a standard spherical halo model fully comprised of 
objects with any mass function inside the range 
$[10^{-7}-4] \; M_{\odot}$.
In the region of stellar mass objects, where this result
improves most on previous ones, the new \lmc\ data contribute
about 60\% to our total sensitivity (the \smc\ and \eros1 
\lmc\ data contribute 15\% and 25\% respectively).  
The total sensitivity, that is proportional to the sum of 
$N_* \, T_{\rm obs} \, \epsilon (t_E)$
over the four \eros\ data sets,
is 2.4 times larger than that of Alcock et al. (1997a).
We observe that a large fraction of the domain previously allowed by 
Alcock et al. (1997a)
is excluded by the limit in Fig.~\ref{excl}.

\section{Discussion and conclusion}
        After eight years of monitoring the Magellanic Clouds, 
\eros\ has a meager crop of three microlensing candidates towards 
the \lmc\ and one towards the \smc , whereas 27 events are expected
for a spherical halo fully comprised of $0.5 \:{\rm M}_\odot$ objects.
These were
obtained from four different data sets analysed by four independent,
cross-validated programs. 
So, the small number of observed events is unlikely
to be due to bad detection efficiencies.

This allows us to put strong constraints on the fraction
of the halo made of objects in the range [$10^{-7}\:{\rm M}_\odot$,
$4\:{\rm M}_{\odot}$], excluding in particular at the 95~\% C.L. that more
than 40~\% of the standard halo be made of objects with up to
$1 \:{\rm M}_\odot$.  The preferred value quoted in Alcock et al. (1997a),
$f = 0.5$ and $0.5\:{\rm M}_\odot$,
is incompatible with the limits in Fig.~\ref{excl} at the 99.7\% C.L.
(but see the note added below).

What are possible reasons for such a difference?
Apart from a potential bias in the detection efficiencies,
several differences  
should be kept in mind while comparing the two experiments.
First, \eros\ uses less crowded fields than \macho\ with the result
that blending is relatively unimportant for \eros .
Second, \eros\ covers a larger solid angle (43~deg$^2$ in the \lmc\
and 10~deg$^2$ in the \smc ) than \macho , which monitors primarily
the central 11~deg$^2$ of the \lmc .
The \eros\ rate should thus be less contaminated by self-lensing
that is more common in the central regions~-
the importance of
self-lensing was first stressed 
by Wu (1994) and Sahu (1994).
Third, the \macho\ data have a more frequent time sampling.
Finally, while the \eros\ limit uses both Clouds, the \macho\
result is based only on the \lmc .
For halo lensing, the timescales towards the two Clouds should
be nearly identical and the optical depths comparable.
In this regard, we remark that the \smc\ event
is longer than all \lmc\ candidates from \macho\ and \eros .

Finally,
given the scarcity of our candidates 
and the possibility that some observed microlenses
actually lie in the Magellanic Clouds, 
\eros\ is not willing to quote at present a non zero 
{\it lower} limit on the fraction of the Galactic halo comprised of
dark compact objects with masses up to a few solar masses.

\smallskip
{\bf Note added.}  While the writing of
this letter was being finalised, the analysis
of 5.7 yrs of \lmc\ observations by the \macho\ group was made
public (\cite{alc00}).  The new favoured estimate of the 
halo mass fraction in the form of compact objects, $f = 0.20$,
is 2.5 times lower than that of Alcock et al. (1997a)
and is compatible with the limit presented here.  
None of the conclusions in this article have to be reconsidered.
A detailed comparison of our results with those of
Alcock et al. (2000) will be available 
in our forthcoming publication (\cite{las00a}).

\begin{acknowledgements}
We are grateful to D. Lacroix and the staff at the Observatoire de
Haute Provence and to A. Baranne for their help with the MARLY
telescope.  
The support by the technical staff at ESO, La Silla, 
is essential to our project.
We thank J.F. Lecointe for assistance with the online computing.
\end{acknowledgements}

\end{document}